\documentclass[12pt]{article}
\usepackage{amsmath,amssymb}

\makeatletter \@addtoreset{equation}{section}
\makeatother

\def\ben{\begin{equation}}
\def\een{\end{equation}}

\let\a=\alpha    
    
  \let\n=\nu

\let\C=\Chi

\def\nn{\nonumber} \def\bd{\begin{document}} \def\ed{\end{document}}
\def\ds{\documentstyle} \let\fr=\frac \let\bl=\bigl \let\br=\bigr
\let\Br=\Bigr \let\Bl=\Bigl
\let\bm=\bibitem
\let\na=\nabla
\let\pa=\partial \let\ov=\overline
\newcommand{\be}{\begin{equation}}
\newcommand{\ee}{\end{equation}}
\def\ba{\begin{array}}
\def\ea{\end{array}}
\def\ft#1#2{{\textstyle{\frac{\scriptstyle #1}{\scriptstyle #2}}}}
\def\fft#1#2{\frac{#1}{#2}}
\def\del{\partial}
\def\vp{\varphi}
\def\sst#1{{\scriptscriptstyle #1}}
\def\oneone{\rlap 1\mkern4mu{\rm l}}
\def\td{\tilde}
\def\wtd{\widetilde}
\def\ie{\rm i.e.\ }
\def\dalemb#1#2{{\vbox{\hrule height .#2pt
        \hbox{\vrule width.#2pt height#1pt \kern#1pt
                \vrule width.#2pt}
        \hrule height.#2pt}}}
\def\square{\mathord{\dalemb{6.8}{7}\hbox{\hskip1pt}}}
\newcommand{\ho}[1]{$\, ^{#1}$}
\newcommand{\hoch}[1]{$\, ^{#1}$}
\newcommand{\bea}{\begin{eqnarray}}
\newcommand{\eea}{\end{eqnarray}}
\newcommand{\ra}{\rightarrow}
\newcommand{\lra}{\longrightarrow}
\newcommand{\Lra}{\Leftrightarrow}
\newcommand{\ap}{\alpha^\prime}
\newcommand{\bp}{\tilde \beta^\prime}
\newcommand{\tr}{{\rm tr} }
\newcommand{\Tr}{{\rm Tr} }
\def\0{{\sst{(0)}}}
\def\1{{\sst{(1)}}}
\def\2{{\sst{(2)}}}
\def\3{{\sst{(3)}}}
\def\4{{\sst{(4)}}}
\def\5{{\sst{(5)}}}
\def\6{{\sst{(6)}}}
\def\7{{\sst{(7)}}}
\def\8{{\sst{(8)}}}
\def\n{{\sst{(n)}}}
\def\cA{{{\cal A}}}
\def\cB{{{\cal B}}}
\def\cF{{{\cal F}}}
\def\tV{\widetilde V}
\def\tW{\widetilde W}
\def\tH{\widetilde H}
\def\tE{\widetilde E}
\def\tF{\widetilde F}
\def\tA{\widetilde A}
\def\im{{{\rm i}}}
\def\tY{{{\wtd Y}}}
\def\ep{{\epsilon}}
\def\vep{{\varepsilon}}
\def\R{\rlap{\rm I}\mkern3mu{\rm R}}
\def\bD{{{\bar D}}}

\def\R{\rlap{\rm I}\mkern3mu{\rm R}}
\def\bD{{{\bar D}}}
\def\R{{{\mathbb R}}}
\def\C{{{\mathbb C}}}
\def\H{{{\mathbb H}}}
\def\CP{{{\mathbb C}{\mathbb P}}}
\def\RP{{{\mathbb R}{\mathbb P}}}
\def\Z{{{\mathbb Z}}}
\def\bA{{{\mathbb A}}}
\def\bB{{{\mathbb B}}}
\def\bC{{{\mathbb C}}}
\def\bD{{{\mathbb D}}}
\def\bE{{{\mathbb E}}}
\def\bZ{{{\mathbb Z}}}
\def\Re{{{\frak{Re}}}}
\def\Im{{{\frak{Im}}}}
\def\cosec{{\,\hbox{cosec}\,}}
\def\Gm{{\Gamma_{\!\! -}}}
\def\Gp{{\Gamma_{\!\! +}}}
\def\stan{{standard }}
\def\nonstan{{supernumerary }}

\newcommand{\auth}{H. L\"u\hoch{\dagger1} and
Justin F. V\'azquez-Poritz\hoch{\ddagger2}}

\begin{document}
\begin{flushright}

MIFP-04-06\ \ \ \ \ UK-04-07\ \ \ \ \
UCTP-107-04\\
{\bf hep-th/0403248}\\
March\  2004
\end{flushright}

%\vspace{10pt}

\begin{center}

{\large {\bf Non-singular Twisted S-branes From Rotating Branes}}

\vspace{20pt}
\auth

\vspace{20pt} {\hoch{\dagger}\it George P. and Cynthia W. Mitchell
Institute for Fundamental Physics,\\ Texas A\& M University,
College Station, TX 77843-4242, USA}

\vspace{10pt} {\hoch{\ddagger}\it Department of Physics and Astronomy,\\
University of Kentucky, Lexington, KY 40506}

\vspace{10pt} {\hoch{\ddagger}\it Department of Physics,\\
University of Cincinnati, Cincinnati OH 45221-0011}

\vspace{30pt}

\underline{ABSTRACT}
\end{center}

We show that rotating $p$-brane solutions admit an analytical
continuation to become twisted S$p$-branes. Although a rotating
$p$-brane has a naked singularity for large angular momenta, the
corresponding S-brane configuration is regular everywhere and
exhibits a smooth bounce between two phases of Minkowski
spacetime. If the foliating hyperbolic space of the transverse
space is of even dimension, such as for the twisted SM5-brane,
then for an appropriate choice of parameters the solution smoothly
flows from a warped product of two-dimensional de Sitter
spacetime, five-dimensional Euclidean space and a hyperbolic
4-space in the infinite past to Minkowski spacetime in the
infinite future. We also show that non-singular S-Kerr solutions
can arise from higher-dimensional Kerr black holes, so long as all
(all but one) angular momenta are non-vanishing for even (odd)
dimensions.

{\vfill\leftline{}\vfill \vskip 10pt \footnoterule {\footnotesize
\hoch{1} Research supported in part by DOE grant
DE-FG03-95ER40917.

{\footnotesize \hoch{2} Research supported in part by DOE grant
DE-FG01-00ER45832.}} \vskip -12pt}  \pagebreak
\setcounter{page}{1}

%\tableofcontents
%\addtocontents{toc}{\protect\setcounter{tocdepth}{2}}
\newpage

\section{Introduction}

     There are two ways of generalizing isotropic extremal $p$-branes
while maintaining the isotropy.  One is to introduce a non-extremal
factor, which gives rise to non-extremal black $p$-branes
\cite{dlpblack,ctblack}. The other is to relax the condition $dA + \td
d B=0$ \cite{lpxtoda}. A nice feature of the second generalization is
that the Poincar\'e symmetry of the $p$-brane world-volume is
maintained. The equations can be reduced to a set of Liouville or Toda
equations which can be solved exactly \cite{lpxtoda}.  These
solutions can be analytically continued \cite{lu2} to give rise to
cosmological solutions \cite{ovrut1,lu1,ovrut2,lu2}, which are now
known as S-branes \cite{gutperle}.

         An interesting feature of the analytical continuation of
static black hole or $p$-brane solutions is that, in many cases,
singular solutions become regular. An early example of this is the
observation that a Schwarzschild black hole can be analytically
continued to a completely regular time-dependent ``bubble of
nothing'' \cite{gott,witten}. Another early example is that a
four-dimensional de Sitter Kerr black hole can give rise to a
smooth four-dimensional Einstein space \cite{4deinstein}. This has
recently been generalized to five dimensions, which results in an
infinite number of five-dimensional inhomogeneous Einstein spaces
\cite{5deinstein}.

It was shown in \cite{adsbh} that singular extremal and
non-extremal charged AdS black holes can be analytically continued
to regular flux-branes. Also, singular de Sitter charged black
holes can be analytically continued to smooth cosmological
solutions \cite{adsbh,desitter}. The cosmological solution that
results in the case of four dimensions can be lifted to a
non-singular S-brane configuration in eleven dimensions
\cite{hong1,hong2}. Recent examples of non-singular S-branes have
also arisen from diholes \cite{strominger} and Kerr black holes
\cite{wang,quevedo}. In this paper, we consider non-singular
S-brane configurations which arise from analytically continuing
rotating $p$-branes. For simplicity of terminology, we shall refer
to these new solutions as S$p$-branes. We also obtain regular
S-Kerr solutions in arbitrary dimenensions.

This paper is organized as follows. In sections 2, 3 and 4, we
present non-singular SM5, SM2 and SD3-branes, which are related to
rotating M5, M2 and D3-branes, respectively. In section 5, we
discuss S-Kerr solutions, which arise from higher-dimensional Kerr
black holes. In section 6, we present general S$p$-branes, which
are related to rotating $p$-branes. Lastly, conclusions are
presented in section 7.

\section{Non-singular SM5-brane}

The rotating M5-brane was constructed in \cite{m51}. Adopting the
notation of \cite{tenauthor}, the metric can be written as
%%%%
\bea ds_{11}^2 &=& H^{-1/3}\Big(
-\Bigl(1-\fft{2m}{r^3\,\Delta}\Bigr)dt^2+dx_1^2+\cdots +dx_5^2\Big)
\nn\\
&& +H^{2/3}\Big[ \fft{\Delta\,dr^2}{H_1\,H_2-\fft{2m}{r^3}}+r^2
\Big( d\mu_0^2+\sum_{i=1}^2 H_i\,(d\mu_i^2+\mu_i^2\,
d\phi_i^2)\Big)\nn\\ &&
-\fft{4m\,\cosh\alpha}{r^3\,H\,\Delta}\,dt\,\sum_{i=1}^2
\ell_i\,\mu_i^2\,d\phi_i+\fft{2m}{r^3\,H\,\Delta}\,(\sum_{i=1}^2
\ell_i\,\mu_i^2\,d\phi_i)^2\Big]\,, \eea
%%%%
where $\Delta$, $H$ and $H_i$ are given by
%%%%
\be \Delta = H_1\,H_2\,\Bigl(\mu_0^2+\fft{\mu_1^2}{H_1}
+\fft{\mu_2^2}{H_2}\Bigr)\,,\qquad
H=1+\fft{2m\,\sinh^2\alpha}{r^3\, \Delta}\,,\qquad
H_i=1+\fft{\ell_i^2}{r^2}\,. \ee
%%%%
The $\mu_i$ satisfy $\mu_0^2+\mu_1^2+\mu_2^2=1$. The 4-form field
strength is given by $F_\4=\ast dA_\6$, where
%%%%
\be
A_\6=\fft{1-H^{-1}}{\sinh\alpha}\,(\cosh\alpha\,dt+\sum_{i=1}^2
\ell_i\,\mu_i^2\, d\phi_i)\wedge d^5x\,. \ee
%%%%
Note that, if one turns off the M5-brane charge by setting
$\alpha=0$, the metric becomes the product of the six-dimensional
Kerr solution with five-dimensional Euclidean space.

     We now perform the following analytical continuation:
%%%%
\be t\rightarrow {\rm i}\,z\,,\qquad r\rightarrow {\rm
i}\,t\,,\qquad \ell_i \rightarrow -{\rm i}\,\ell_i\,,\qquad \alpha
\rightarrow {\rm i}\,\alpha\,,\qquad m\rightarrow -{\rm i}\,m\,,
\ee
%%%%
and define
%%%%
\be \mu_0=\tilde \mu_0\,,\qquad \mu_1={\rm i}\,\tilde
\mu_1\,,\qquad \mu_2={\rm i}\,\tilde \mu_2\,, \ee
%%%%
such that $\tilde \mu_0^2-\tilde \mu_1^2-\tilde \mu_2^2=1$.  It is
important to note that the $\td \mu_i$ are unbounded now. The rotating
M5-brane becomes a twisted SM5-brane, given by
%%%%
\bea ds_{11}^2 &=& \tilde H^{-1/3}\Big(
\Bigl(1-\fft{2m}{t^3\,\tilde \Delta}\Bigr)dz^2+dx_1^2+\cdots
+dx_5^2\Big)\nn\\
&& +\tilde H^{2/3}\Big[ -\fft{\tilde \Delta\,dt^2}{\tilde
H_1\,\tilde H_2-\fft{2m}{t^3}}+ t^2 \Big( -d\tilde
\mu_0^2+\sum_{i=1}^2 \tilde H_i\,(d\tilde \mu_i^2+\tilde \mu_i^2\,
d\phi_i^2)\Big)\nn\\ && +\fft{4m\,\cos\alpha}{t^3\,\tilde
H\,\tilde \Delta}\,dz\,\sum_{i=1}^2 \ell_i\,\tilde
\mu_i^2\,d\phi_i-\fft{2m}{t^3\,\tilde H\,\tilde
\Delta}\,(\sum_{i=1}^2
\ell_i\,\tilde \mu_i^2\,d\phi_i)^2\Big]\nn\,,\\
A_\6 &=& \fft{1-\tilde
H^{-1}}{\sin\alpha}\,(\cos\alpha\,dz+\sum_{i=1}^2 \ell_i\,\tilde
\mu_i^2\, d\phi_i)\wedge d^5x\,, \eea
%%%%
where
%%%%
\be \tilde H=1-\fft{2m\,\sin^2\alpha}{t^3\,\tilde \Delta}\,,\qquad
\tilde H_i=1+ \fft{\ell_i^2}{t^2}\,,\qquad \tilde \Delta=\tilde
H_1\,\tilde H_2\,\Bigl( \tilde \mu_0^2 -\fft{\tilde
\mu_1^2}{\tilde H_1}-\fft{\tilde \mu_2^2}{\tilde H_2}\Bigr) \,.
\ee
%%%%
Notice that the solution is invariant under $t\rightarrow -t$,
$m\rightarrow -m$.

Unlike the corresponding M5-brane, this S-brane solution does not
have an extremal limit. The brane charge vanishes when $m=0$, in
which case the metric is merely that of Minkowski spacetime. Thus,
for non-triviality, we consider the case with $m\ne 0$.  Another
difference between this S-brane and the corresponding M5-brane is
that, for the latter solution, one can perform a decoupling limit
in which the ``1'' in the function $H$ can be dropped.  The
resulting solution can be consistently dimensionally reduced to a
seven-dimensional AdS black hole \cite{tenauthor}. However, this
procedure is not possible for the SM5-brane, for the same reason
that an extremal limit is lacking. This feature persists in all of
the S$p$-branes that we have obtained. This implies that these
S$p$-branes are intrinsically higher-dimensional and do not arise
from the dimensional oxidation of lower-dimensional gauged
supergravity solutions.

     When all $\ell_i=0$, the solution is an analytical continuation
of the black M5-brane. This is different from the standard
S5-brane \cite{lu2,gutperle}, since in the present case the
Euclidean symmetry of the worldvolume is broken. Instead, standard
S-branes are related by analytical continuation to non-extremal
$p$-branes which have Poincar\'e symmetry on the worldvolume
\cite{lu2}. Also, note that the presence of the $\ell_i$
parameters breaks the R-symmetry associated with the isometries of
the transverse hyperbolic space. The same story holds for all of
the S-brane solutions presented in this paper.

      Now let us study the behavior of the metric.  In the regions
$t\rightarrow \pm\infty$, the metric becomes Minkowski spacetime
written in the following form:
%%%%%
\be ds_{11}^2 = -dt^2 + t^2 d\Omega_{4,-1}^2 + dz^2 + dx_1^2 +
\cdots + dx_5^2\,, \ee
%%%
where $d\Omega_{4,-1}^2=-d\td\mu_0^2 + \sum_{i=1}^2 (d\td\mu_i^2 +
\td\mu_i^2\,d\phi_i^2)$ is the metric of the unit hyperbolic
4-plane. This describes a five-dimensional universe with hyperbolic
spatial slices expanding or contracting at a constant rate, together
with a stable $\bE ^5$ or $T^5$. To analyze the behavior at small $t$,
we first examine the $dt^2$ term.  Since the quantity
%%%
\be t^4 \td\Delta = (t^2 + \ell_1^2)(t^2 + \ell_2^2) \Bigl(1 +
\fft{\td\mu_1^2\,\ell_1^2}{t^2 + \ell_1^2} +
\fft{\td\mu_2^2\,\ell_2^2}{t^2 + \ell_2^2}\Bigr)\,,\label{t4delta}
\ee
%%%%
is positive definite for all $\ell_i$ non-vanishing , we can
determine whether $t$ is a time-like or space-like coordinate by
examining the function $F$, given by
\be F=t^4 (\td H_1\, \td H_2) -2m t\,. \ee
%%%
This function is positive for $t\rightarrow \pm \infty$. If
$|m|<m_0$, with $m_0$ given by
%%%
\be m_0 =\fft{\sqrt6}{18}\, (\sqrt{\ell_1^4 +14\ell_1\ell_2
+\ell_2^4} +2\ell_1^2 + 2\ell_2^2)\, \sqrt{\sqrt{\ell_1^4
+14\ell_1\ell_2 +\ell_2^4} - \ell_1^2 -\ell_2^2}\,, \label{m0} \ee
%%%
then the function $F$ is positive definite and $t$ is always
time-like.  On the other hand, if $|m|> m_0$, then the function
has two positive roots $t_+$ and $t_-$. In the regions $t>t_+$ and
$t<t_-$, the function $F$ is positive and $t$ is time-like. In the
region $t_-< t< t_+$, $F$ is negative and $t$ becomes a space-like
coordinate. When $|m|=m_0$, the function $F$ has a second-order
zero. In other words, we have $t_+=t_-=t_0$, where $t_0$ is given
by
%%%%
\be
t_0 = \fft{1}{\sqrt6}\, \sqrt{\sqrt{\ell_1^4 +14\ell_1\ell_2
+\ell_2^4} - \ell_1^2 -\ell_2^2}\,.\label{mt05b}
\ee
%%%%%%

     Let us first consider the case for which $|m|< m_0$. The twisted
SM2-brane is then completely regular, bouncing from one Minkowski
spacetime in the infinite past to another Minkowski spacetime in
the infinite future. To show this, we note that the function $\td H$,
given by
%%%%
\bea
\td H=\fft{1}{t^4\, \td\Delta}\, \Bigl( F +2m\, t\, \cos^2\alpha
+t^4 \td H_1\, \td H_2 \sum_{i=1}^2 \fft{\td\mu_1^2\,\ell_i^2}{
t^2 +\ell_i^2}\Bigr)\,,\label{sm5h}
\eea
%%%%
is positive definite if $F$ is non-negative.  One might worry that the
$2mt\,\cos^2\a$ term may cause a problem for negative $t$.  However,
$F+2mt\,\cos^2\alpha=t^4\td H_1\,\td H_2-2mt\,\sin^2\alpha$ is always
positive if $F$ is non-negative.
Thus, we find that $0<\td H\le 1$.  This implies that the
scale factor for the five-dimensional Euclidean space $dx^i\, dx^i$
and $dz_i^2$ remains finite and non-vanishing, while the hyperbolic
4-plane undergoes contraction and expansion.

      One important difference between the twisted SM5-brane and
the original rotating M5-brane is that $t^4\td\Delta$ is positive
definite for the former case as long as all $\ell_i$ are
non-vanishing.  The corresponding term $r^4\Delta$ for the
rotating M5-brane always encounters a singularity.  This is the
reason why the rotating M5-brane with two large angular momentum
has a naked singularity while the corresponding SM5-brane is
regular.

       When $|m|>m_0$, the function $\td H$ may become zero for large
$|m|$.  This happens when $m\ge m_0/\sin^2\alpha$.  The solution
then has a naked singularity when $\td H$ vanishes. For $m_0<|m|<
m_0/\sin^2\alpha$, although $F$ can be negative, $\td H$ is still
positive definite.  The solution becomes stationary in the region
$t_-<t<t_+$, where $t_\pm$ are the horizons.  It is expected that
the $dz^2$ term can change sign in this region, and $z$ is then
promoted as a time coordinate. However, the periodic coordinates
$\phi_i$ can also develop a time-like signature.  To see this, we
note that the determinant of the metric in the $z$ and $\phi_i$
directions is given by
%%%
\be \det(g_{ij}(d\phi_i, dz)) = \td H\,\td \mu_1^2\,\td \mu_2^2\,
F\,. \ee
%%%%
We see that, when $F<0$, the signature of either $z$ or one of the
$\phi_i$ is changed.  Thus, in the region where $F<0$ but with
positive $(1-2m/(t^3\,\td\Delta))$, one of the compact coordinates
$\phi_i$ will become time-like.  To see this more precisely, let
us examine, without loss of generality, the $d\phi_1^2$ component
of the metric; it is given by
%%%%%
\be \fft{t^2\td H_1\, (F + 2mt\, \cos^2\alpha) +
\td\mu_1^2\,\ell_1^2\, F + t^4\td H_1\,\td\mu_2^2\,\ell_2^2}{
t^4\, \td \Delta\td H}\,\td \mu_1^2\,. \ee
%%%%
We see that, for $F>0$, it is positive definite.  When $F$ becomes
negative then, since $\td\mu_1^2$ is unbounded, the compact
coordinate $\phi_1$ can become time-like, giving rise to a closed
time-like curve. Since this stationary region is not hidden from
external observers by an event horizon, it is not clear if this
case has physical relevance.

      Now let us consider the case when $|m|=m_0$, for which
%%%
\be
F(t_0)=0\,,\qquad F'(t_0)=0\,.
\ee
%%%
In other words, there exists a second-order zero for the function
$F$ at $t_0$.  The function $F$, however, never actually becomes
negative. In this case, the Minkowski spacetimes at $t\rightarrow
\pm\infty$ are disjoint. The point of $t=t_0$ can be viewed as a
two-dimensional de Sitter spacetime in the infinite past. To see
this, we complete the square for the $d\phi_i$ terms, and the
remaining $dz^2$ metric component is given by
%%%
\be
\fft{(\td H_1\,\td H_2\, t^3 - 2m)\, \td\Delta\, \td
H^{\fft23}}{ \td \Delta (t^3 \td H_1\, \td H_2 -2m) + 2\td
H_1\,\td H_2\, m\,\cos^2\alpha}\,,
\ee
%%%%%%
which shrinks to zero at $t=t_0$.  In the $t=t_0$ region, the
metric for the $t$ and $z$ coordinates becomes
%%%%
\be ds^2 = t_0^4\,\td\Delta\, \td H^{\ft23}\,\Bigl(
\fft{(t-t_0)^2\, F''(t_0)}{4t_0^5\,\td H_1\, \td H_2\, m\,
\cos\alpha}\, dz^2 - \fft{2}{(t-t_0)^2\, F''(t_0)}\,dt^2 \Bigr)\,,
\ee
%%%
which is two-dimensional de Sitter spacetime with a warp factor
$\td \Delta\, \td H^{2/3}$ that depends on the coordinates of the
four-dimensional hyperbolic space.  In the special case
$\ell_1=\ell_2\equiv \ell$, the conditions (\ref{m0}) and
(\ref{mt05b}) reduce to
%%%
\be m_0=\fft{8\ell^3}{3\sqrt3}\,,\qquad t_0=\fft{\ell}{\sqrt3}\,,
\ee
%%%
and the function $F$ simplifies to
%%%
\be F=(t-t_0)^2 (t + 2t\, t_0 + 9 t_0^2)\,. \ee
%%%

         To summarize, for all $\ell_i$ non-vanishing, $m$ has a
critical value $m_0$.  For $|m|<m_0$, the solution is regular
everywhere and bounces between two phases of five-dimensional
Minkowski spacetime, with a stable $\bE ^6$ or $T^6$. If $m=m_0$,
then the geometry runs from a warped product of two-dimensional de
Sitter spacetime and a hyperbolic 4-plane with a stable $\bE ^5$
or $T^5$ to Minkowski spacetime. For $|m|> m_0$, the solution
develops closed time-like curves in the stationary region. The
solution is further corrupted by a naked singularity when $m\ge
m_0/\sin^2\alpha$, due to the contribution of the brane charge.

           So far, we have considered the case when both $\ell_i$ are
non-zero.  When one or both of them vanish, the solution becomes
singular.  This is because $F$ will no longer be positive definite
and, furthermore, the condition $F(t_0)=0$ and $F'(t_0)=0$ can
never be satisfied.  Thus, there is nothing to prevent the
solution passing from the time-like region to the stationary
region, where the $t^4\td\Delta$ term in (\ref{t4delta}) becomes
zero at $t=0$. This yields a naked singularity in the stationary
region.

      There can be other analytical continuations than what has been
presented above.  For example, we can take $\mu_0={\rm i}\, \td
\mu_0$, $\mu_1={\rm i}\, \td \mu_1$ and $\mu_2={\rm i}\, \td
\mu_2$. The reality of the solution now requires $\phi_i$, rather
than $\ell_i$ to undergo a Wick rotation. As a result, the
functions $\td H_i$ are given by $\td H_i=1-\ell_i^2/t^2$. It is
easy to verify that the solution is not regular.

\section{Non-singular SM2-brane}

The metric for the rotating M2-brane was obtained in \cite{m51}.
Adopting the notation of \cite{tenauthor}, the solution is given
by
%%%%
\bea ds_{11}^2 &=& H^{-2/3}\Big(
-\Bigl(1-\fft{2m}{r^6\,\Delta}\Bigr)dt^2+dx_1^2+dx_2^2\Big)\nn\\
&& +H^{1/3}\Big[
\fft{\Delta\,dr^2}{H_1\,H_2\,H_3\,H_4-\fft{2m}{r^6}}+r^2
\sum_{i=1}^4 H_i\,(d\mu_i^2+\mu_i^2\, d\phi_i^2)\nn\\
&& -\fft{4m\,\cosh\alpha}{r^6\,H\,\Delta}\,dt\,(\sum_{i=1}^4
\ell_i\,\mu_i^2\,d\phi_i)+\fft{2m}{r^6\,H\,\Delta}\,(\sum_{i=1}^4
\ell_i\,\mu_i^2\,d\phi_i)^2\Big]\,, \eea
%%%%
where $\Delta$, $H$ and $H_i$ are given by
%%%%
\be \Delta = H_1\,H_2\,H_3\,H_4\,\sum_{i=1}^4
\fft{\mu_i^2}{H_i}\,,\qquad H=1+\fft{2m\,\sinh^2\alpha}{r^6\,
\Delta}\,,\qquad H_i=1+\fft{\ell_i^2}{r^2}\,, \ee
%%%%
and the $\mu_i$ satisfy $\sum_{i=1}^4\mu_i^2=1$. The 3-form gauge
potential is given by
%%%%
\be
A_\3=\fft{1-H^{-1}}{\sinh\alpha}\,(-\cosh\alpha\,dt+\sum_{i=1}^4
\ell_i\,\mu_i^2\, d\phi_i)\wedge d^2x\,. \ee
%%%%
We perform the following analytical continuation:
%%%%
\bea && t\rightarrow {\rm i}\,z\,,\qquad r\rightarrow {\rm
i}\,t\,,\qquad
\phi_4 \rightarrow {\rm i}\,\phi_4\,,\nn\\
&& \alpha \rightarrow {\rm i}\,\alpha\,,\qquad m\rightarrow -{\rm
i}\,m\,,\qquad \ell_j\rightarrow -{\rm i}\,\ell_j\qquad j=1,2,3\,,
\eea
%%%%
and $\ell_4$ is unchanged. We define
%%%%
\be \mu_4=\tilde \mu_4\,,\qquad \mu_j={\rm i}\,\tilde \mu_j \,,
\ee
%%%%
such that $\tilde \mu_4^2-\sum_{j=1}^3 \tilde \mu_j^2=1$.  Thus
the $\td \mu_i$ are unbounded.

The rotating M2-brane becomes a twisted SM2-brane given by
%%%%
\bea ds_{11}^2 &=& \tilde H^{-2/3}\Big(
\Bigl(1-\fft{2m}{t^6\,\tilde \Delta}\Bigr)\,dz^2+dx_1^2+dx_2^2\Big)
\nn\\
&& +\tilde H^{1/3}\Big[ -\fft{\tilde \Delta\,dt^2}{\tilde
H_1\,\tilde H_2\,\tilde H_3\,\tilde H_4-\fft{2m}{t^6}}+
t^2\sum_{j=1}^3 \tilde H_j\,(d\tilde \mu_j^2+\tilde \mu_j^2\,
d\phi_j^2)\nn\\
&&+t^2\, \tilde H_4\,(-d\tilde \mu_4^2+\tilde
\mu_4^2\,d\phi_4^2)+\fft{4m\,\cos\alpha}{t^6\,\tilde H\,\tilde
\Delta}\,dz\,(\sum_{i=1}^4 \ell_i\,\tilde
\mu_i^2\,d\phi_i)\nn\\
&&-\fft{2m}{t^6\,\tilde H\,\tilde
\Delta}\,(\sum_{i=1}^4 \ell_i\,\tilde
\mu_i^2\,d\phi_i)^2\Big]\,,\nn\\ A_\3 &=& \fft{1-\tilde
H^{-1}}{\sin\alpha}\,(-\cos\alpha\,dz+\sum_{i=1}^4 \ell_i\,\tilde
\mu_i^2\, d\phi_i)\wedge d^2x\,, \eea
%%%%
where
%%%%
\bea \tilde H &=& 1-\fft{2m\,\sin^2\alpha}{t^6\,\tilde
\Delta}\,,\qquad \tilde \Delta=\tilde H_1\,\tilde H_2\,\tilde
H_3\,\tilde H_4\,\Bigl(\fft{\tilde \mu_4^2}{\tilde H_4}
-\sum_{j=1}^3\fft{\tilde \mu_j^2}{\tilde
H_j}\Bigr)\,,\nn\\
H_4 &=& 1-\fft{\ell_4^2}{t^2}\,,\qquad \tilde H_j=1+
\fft{\ell_j^2}{t^2}\qquad j=1,2,3\,. \eea
%%%%
This solution is invariant under $t\rightarrow -t$.

      In the asymptotic $t\rightarrow \pm\infty$ regions, the geometry
becomes Minkowski spacetime, describing an eight-dimensional
universe expanding or contracting at a constant rate, together
with a stable $\bE^3$ or $T^3$. At small $t$, the analysis is
similar to the previous section. In order to avoid a singularity
when $H_4=1-\ell_i/t^2$ vanishes or becomes negative, it is
necessary to set $\ell_4=0$.   Having done that, the quantity
%%%%
\be
t^6\td \Delta = (t^2 +\ell_1)^2(t^2 + \ell_2^2)(t^2 + \ell_3^2)
\Bigl(1 + \sum_{i=1}^3 \fft{\td\mu_i^2\,\ell_i^2}{t^2 + \ell_i^2}\Bigl)
\ee
%%%%
is positive definite as long as $\ell_1$, $\ell_2$ and $\ell_3$ are
all non-vanishing.  The global structure of the metric is now largely
determined by the function
%%%%
\be
F=t^6\td H_1\, \td H_2\, \td H_3 - 2m\,.
\ee
%%%%
When we have
%%%
\be
m<m_0\equiv \ft12 (\ell_1\,\ell_2\,\ell_3)^2\,,\qquad
\ell_4=0\,,
\ee
%%%%
the function $F$ is positive definite.  The function $\td H$, given
by
%%%
\be
\td H=t^{-6}\td \Delta^{-1}\Bigl( F + 2m\cos^2\alpha +
t^6\td H_1\,\td H_2\,\td H_3\sum_{i=1}^3
\fft{\td\mu_i^2\,\ell_i^2}{t^2 + \ell_i^2}\Bigr)\,,
\ee
%%%%
can also be seen to be positive definite.  Thus, the solution is
regular everywhere, bouncing between two Minkowski spacetimes in
the $t\rightarrow \pm\infty$ regions.  It is worth pointing out
that the condition $m<m_0$ includes the possibility of negative
$m$.

     When $m\ge m_0$, the function $F$ has two real roots
$\pm t_0$, between which the solution becomes stationary.  For
$m\ge m_0/\sin^2\alpha$, the function $\td H$ can vanish and,
hence, the solution is singular.  For $m_0\le m <
m_0/\sin^2\alpha$, we expect that the solution has no curvature
singularity. However, metric develops closed time-like curves.
This can be seen from the determinant of the metric for the $z$
and $\phi_i$ directions, which is given by
%%%
\be \det(g_{ij})=\td H^{2/3}\,\td\mu_1^2\,\td\mu_2^2\,\td\mu_3^2\,
\td\mu_4^2\, F\,. \ee
%%%
Comparing this with the $dz^2$ term, it is easy to see that the
compact coordinates $\phi_i$ can develop time-like signatures.
Without loss of generality, we can examine the $d\phi_1^2$ term,
given by
%%%%
\be \fft{\td\mu_1^2\,d\phi_1^2}{t^6\,\td H\, \td \Delta}\,\Bigl[
t^2\td H_1\, (F +2m\cos^2\alpha) + F\,\td\mu_1^2\,\ell_1^2 + t^6
\td H_1\,\td H_2\,\td H_3\sum_{i=2}^3
\fft{\td\mu_i^2\,\ell_i^2}{t^2\td H_i}\Bigr]\,. \ee
%%%%%
The positivity of this term is ensured only for positive $F$,
since $\td\mu_i$ is unbounded.

      To summarize, when $\ell_4=0$ and $\ell_j\ne 0$ with $j=1,2,3$,
there exists a critical value $m_0=\ft12
(\ell_1\,\ell_2\,\ell_3)^2$.  For $-\infty<m<m_0$, the solution is
regular everywhere, bouncing between two Minkowski spacetimes at
$t\rightarrow \pm \infty$. The solution is totally symmetric under
$t\rightarrow -t$. The function $\td H$ is finite but
non-vanishing, which implies the stability of a $\bE^3$ (or $T^3$)
portion of the spacetime. For $m>m_0$, the solution develops
closed time-like curves. Furthermore, for $m>m_0/\sin^2\alpha$ the
solution becomes singular, due to the brane charge contribution.
It is worth pointing out that unlike the SM5-brane, here we are forced
to set one of the angular momentum , which implies that $F$ never develops
a second-order zero.  Thus there can be no dS$_2$ factor arising in this
case.  This turns out to be a generic feature of twisted S-branes with
an odd-dimensional hyperbolic transverse space.

         If there are additional $\ell_i$ vanishing, the solution
becomes singular at $t=0$, which is in a stationary region.  There
are alternative analytical continuations which lead to singular
solutions.

\section{Non-singular SD3-brane}

The metric for the rotating D3-brane was found in
\cite{kraus,russo}. Adopting the notation of \cite{tenauthor}, the
solution is given by
%%%%
\bea ds_{10}^2 &=& H^{-1/2}\Big(
-\Bigl(1-\fft{2m}{r^4\,\Delta}\Bigr)dt^2+dx_1^2+dx_2^2+dx_3^2\Big)\nn\\
&& +H^{1/2}\Big[
\fft{\Delta\,dr^2}{H_1\,H_2\,H_3-\fft{2m}{r^4}}+r^2
\sum_{i=1}^3 H_i\,(d\mu_i^2+\mu_i^2\, d\phi_i^2)\nn\\
&& -\fft{4m\,\cosh\alpha}{r^4\,H\,\Delta}\,dt\,\sum_{i=1}^3
\ell_i\,\mu_i^2\,d\phi_i+\fft{2m}{r^4\,H\,\Delta}\,(\sum_{i=1}^3
\ell_i\,\mu_i^2\,d\phi_i)^2\Big]\,, \eea
%%%%
where $\Delta$, $H$ and $H_i$ are given by
%%%%
\be \Delta = H_1\,H_2\,H_3\,\sum_{i=1}^3
\fft{\mu_i^2}{H_i}\,,\qquad H=1+\fft{2m\,\sinh^2\alpha}{r^4\,
\Delta}\,,\qquad H_i=1+\fft{\ell_i^2}{r^2}\,, \ee
%%%%
and the $\mu_i$ satisfy $\mu_1^2+\mu_2^2+\mu_3^2=1$. The self-dual
5-form field strength is given by $F_\5=dA_\4+\ast dA_\4$, where
%%%%
\be
A_\4=\fft{1-H^{-1}}{\sinh\alpha}\,(-\cosh\alpha\,dt+\sum_{i=1}^3
\ell_i\,\mu_i^2\, d\phi_i)\wedge d^3x\,. \ee
%%%%
We perform the following analytical continuation:
%%%%
\bea && t\rightarrow {\rm i}\,z\,,\qquad r\rightarrow {\rm
i}\,t\,,\qquad
\phi_3\rightarrow {\rm i}\,\phi_3\,,\nn\\
&& \ell_1 \rightarrow -{\rm i}\,\ell_1\,,\qquad \ell_2 \rightarrow
-{\rm i}\,\ell_2\,,\qquad \alpha \rightarrow {\rm i}\,\alpha\,,
\eea
%%%%
and $\ell_3$ and $m$ are unchanged. We define
%%%%
\be \mu_1={\rm i}\,\tilde \mu_1\,,\qquad \mu_2={\rm i}\,\tilde
\mu_2\,,\qquad \mu_3=\tilde \mu_3\,, \ee
%%%%
such that $\tilde \mu_3^2-\tilde \mu_1^2-\tilde \mu_2^2=1$.

The rotating D3-brane becomes a twisted SD3-brane, given by
%%%%
\bea ds_{10}^2 &=& \tilde H^{-\ft12}\Big(
\Bigl(1-\fft{2m}{t^4\,\tilde \Delta}\Bigr)dz^2+dx_1^2+dx_2^2+dx_3^2\Big)
\nn\\
&& +\tilde H^{\ft12}\Big[ -\fft{\tilde \Delta\,dt^2}{\tilde
H_1\,\tilde H_2\,\tilde H_3-\fft{2m}{t^4}}+t^2 \Big( \sum_{j=1}^2
\tilde H_j\,(d\tilde \mu_j^2+\tilde \mu_j^2\, d\phi_j^2)
-\td H_3(d\tilde \mu_3^2-\tilde \mu_3^2\,d\phi_3^2)\Big)\nn\\ &&
+\fft{4m\,\cos\alpha}{t^4\,\tilde H\,\tilde
\Delta}\,dz\,\sum_{i=1}^3 \ell_i\,\tilde
\mu_i^2\,d\phi_i-\fft{2m}{t^4\,\tilde H\,\tilde
\Delta}\,(\sum_{i=1}^3 \ell_i\,\tilde
\mu_i^2\,d\phi_i)^2\Big]\,,\nn\\
A_\4 &=& \fft{1-\tilde
H^{-1}}{\sin\alpha}\,(-\cos\alpha\,dz+\sum_{i=1}^3 \ell_i\,\tilde
\mu_i^2\, d\phi_i)\wedge d^3x\,, \eea
%%%%
where
%%%%
\bea \tilde H &=& 1-\fft{2m\,\sin^2\alpha}{t^4\,\tilde
\Delta}\,,\qquad \tilde \Delta=\tilde H_1\,\tilde H_2\,\tilde
H_3\,\Bigl(\fft{\tilde \mu_3^2}{\tilde H_3} -\fft{\tilde
\mu_1^2}{\tilde
H_1}-\fft{\tilde \mu_2^2}{\tilde H_2}\Bigr)\,,\nn\\
\td H_3 &=& 1-\fft{\ell_3^2}{t^2}\,,\qquad \tilde H_j=1+
\fft{\ell_j^2}{t^2}\qquad j=1,2\,. \eea
%%%%
This solution is invariant under $t\rightarrow -t$.

       Since the foliating sphere for D3-brane is odd-dimensional, the
properties of the SD3-brane solution are rather analogous to those
of the SM2-brane.  In the asymptotic $t\rightarrow \pm\infty$
regions, the geometry becomes Minkowski spacetime, describing a
six-dimensional expanding or contracting universe, together with a
stable $\bE^4$ or $T^4$.

       For regularity, it is necessary to set $\ell_3$ to zero.  When
the remaining $\ell_1$ and $\ell_2$ are both non-zero. there
exists a critical value $m_0=\ft12 (\ell_1\,\ell_2\,\ell_3)^2$.
For $-\infty<m<m_0$, the solution is regular everywhere, bouncing
between two six-dimensional Minkowski spacetimes at $t\rightarrow
\pm \infty$ with a stable $\bE^4$ (or $T^4$). The solution is
totally symmetric under $t\rightarrow -t$.  The function $\td H$
is finite but non-vanishing, implying the stability of the $\bE^4$
(or $T^4$). For $m>m_0$, the solution develops closed time-like
curves. Furthermore, for $m>m_0/\sin^2\alpha$, the solution
becomes singular due to the brane charge contribution.

         If there are additional $\ell_i$ vanishing, the solution
becomes singular at $t=0$, which is in a stationary region. There
exist more ways of taking analytical continuations, all of which
lead to singular solutions.

\section{S-Kerr solutions}

           So far, we have considered three examples of rotating
$p$-branes analytically continuing to S-brane configurations. The
$S$-brane solutions can become regular for large angular momenta.
One feature is that the introduction of the brane charge does not
affect the regularity of the solutions in any significant way.
Thus, the discussion of the regularity is more or less the same
for S-Kerr solutions, which can be analytically continued from
Kerr black holes. Regular S-Kerr solutions for four and five
dimensions were obtained in \cite{wang,quevedo}.
Higher-dimensional S-Kerr solutions with one angular momentum
parameter were obtained in \cite{quevedo} but these solutions are
singular.  In this section, we obtain regular S-Kerr solutions in
arbitrary dimension.  The situation differs depending on whether
the number of dimensions is even or odd.

\subsection{Even dimensions: $D=2n+2$}

The Kerr-Schild solution in $D=2n+2$ dimensions, with a sphere of
$d=2n$ dimensions, has the metric \cite{mp}
%%%%
\bea ds_D^2&=& - dt^2 + \fft{\Delta\, dr^2}{\prod_{i=1}^n
H_i-\fft{2m}{r^{2n-1}}} + r^2 d\mu_0^2 + r^2\,
\sum_{i=1}^n H_i\, (d\mu_i^2 + \mu_i^2\, d\phi_i^2)\nn\\
&&+\fft{2m}{r^{2n-1}\, \Delta}\, (dt -\sum_{i}^n \ell_i\,
\mu_i^2\,d\phi_i)^2\,,
\eea
%%%%
where
%%%
\be
\mu_0^2 + \sum_{i=1}^n \mu_i^2=1\,,\qquad \Delta =
\Bigl(\mu_0^2 + \sum_{i=1}^n \fft{\mu_i^2}{H_i}\Bigr)\, \prod_{i=1}^n
H_i\,,
\ee
%%%%
and $H_i=1 + \ell_i^2/r^2$.  We can now perform the analytical
continuation
%%%
\be t\rightarrow {\rm i}\,z\,,\qquad r\rightarrow {\rm
i}\,t\,,\qquad \ell_i \rightarrow -{\rm i}\,\ell_i\,,\qquad \alpha
\rightarrow {\rm i}\,\alpha\,,\qquad m\rightarrow -{\rm i}\,m\,,
\ee
%%%%
together with taking $\mu_0=\td\mu_0$ and $\mu_i={\rm i}\, \td
\mu_i$, such that $\td\mu_0^2 - \sum_{i=1}^n \td\mu_i^2=1$. The
Kerr-Schild solution becomes an S-Kerr solution, given by
%%%%%
\bea ds_D^2&=& dz^2 - \fft{\td\Delta\, dt^2}{\prod_{i=1}^n \td
H_i-\fft{2m}{t^{2n-1}}} - t^2 d\td \mu_0^2 + t^2\,
\sum_{i=1}^n \td H_i\, (d\td\mu_i^2 + \td\mu_i^2\, d\phi_i^2)\nn\\
&&-\fft{2m}{t^{2n-1}\, \td\Delta}\, (dz -\sum_{i}^n \ell_i\,
\td\mu_i^2\,d\phi_i)^2\,,\label{skerreven} \eea
%%%%%
where
%%%
\be \td H_i=1 +\fft{\ell_i^2}{t^2}\,,\qquad \td
\Delta=\Bigl(\td\mu_0^2 -\sum_{i=1}^n \fft{\td\mu_i^2}{\td
H_i}\Bigr) \prod_{i=1}^n\td H_i\,. \ee
%%%
This solution is invariant under $t\rightarrow -t$, $m\rightarrow
-m$.
%%%%

     In the asymptotic $t\rightarrow \pm \infty$ regions, the solution
becomes a $D-1$ dimensional universe with hyperbolic spatial
slices expanding or contracting at a constant rate, together with
a stable $\R$ or $S^1$ direction. Note that the quantity
$t^{2n}\td \Delta$\, given by
%%%
\be t^{2n}\, \td\Delta = \Bigl(1 + \sum_{i=1}^n
\fft{\td\mu_i^2\,\ell_i^2}{t^2 + \ell_i^2}\Bigr)\prod_{i=1}^n (t^2 +
\ell_i^2)\,, \ee
%%%%
is positive definite for non-vanishing $\ell_i$.  That is, if any
of the $\ell_i$ is zero, then the above quantity will vanish at
$t=0$ and the solution becomes singular.  The global structure is
largely determined by the function
%%%%
\be
F=-2mt + t^{2n} \prod_{i=1}^n\td H_i\,,
\ee
%%%%
which appears as a factor in the $dt^2$ term of the metric
(\ref{skerreven}).  When all the parameters $\ell_i$ are non-zero,
there exists a critical value $m_0$, for which there is a time
$t_0$ such that $F(t_0)=0$ and $F'(t_0)=0$.  The values of $m_0$
and $t_0$ depend on the $\ell_i$ in a complicated way.  We only
present a special case of $\ell_i=\ell$:
%%%
\be
t_0=\fft{\ell}{\sqrt{2n-1}}\,,\qquad
m_0=2^{n-1}\, n^n\, t_0^{2n-1}\,.
\ee
%%%%

     When $|m|<m_0$, the function $F$ is positive definite. It follows
that the solution is regular everywhere, bouncing between two
Minkowski spacetimes.  When $|m|>m_0$, the function $F$ will have
two roots $t_+$ and $t_-$, between which $F$ becomes negative and
the solution is stationary.  Although we expect that there is no
curvature singularity in this case, the metric develops closed
time-like curves.   This can be seen, without loss of generality,
by examining the $d\phi_1^2$ term of the metric, given by
%%%%
\be \fft{\td\mu_1^2}{t^{2n}\td \Delta}\Bigl[ t^{2n+2}\prod_{i=1}^n
\td H_i + F\,\td\mu_1^2\,\ell_i^2 + t^{2n} (\prod_{i=1}^n \td
H_i)\sum_{i=2}^n \fft{\td \mu_i^2\,\ell_i^2}{t^2\,\td H_i}\Bigr]
\,.\label{tdeltagen} \ee
%%%%%
Thus, the compact coordinates $\phi_i$ are always space-like if
$F$ is positive but can become time-like if $F$ is negative, since
the $\td \mu_i$ are not bounded.

          Finally, when $|m|=m_0$, the function $F$ has a second-order
zero at $t=t_0$ but never becomes negative.  Near $t=t_0$, the
metric contains an element of two-dimensional de Sitter spacetime.
To see this, we can complete the square of $d\phi_i$, and the
remaining $dz^2$ term becomes
%%%%%
\be \fft{\td \Delta\, F}{\td \Delta\, F + 2mt\prod_{i=1}^n \td
H_i}\,. \ee
%%%%
Thus, the metric components involving the $t$ and $z$ coordinates
around $t=t_0$ are given by
%%%%
\be ds^2 = \td\Delta \Bigl[ \fft{(t-t_0)^2\, F''(t_0)}{4m\, t_0
\prod_{i=1}^n \td H_i}\,dz^2 - \fft{2t_0^{2n}}{(t-t_0)^2\,
F''(t_0)}\,dt^2 \Bigr]\,, \ee
%%%%
where $\td H_i=1+\ell_i^2/t_0^2$.  The metric is clearly
two-dimensional de Sitter spacetime with $t=t_0$ being the
infinite past in the comoving frame.  It has a warp factor $\td
\Delta$ that depends on the coordinates of the hyperbolic space.
Thus, the $D$-dimensional metric at $t=t_0$ describes a warp
product of dS$_2$ and a hyperbolic $2n$-space.

     To summarize, the S-Kerr solution in even-dimensions is regular,
provided that all the $\ell_i$ parameters are turned on and
$|m|\le m_0$.  For $|m|<m_0$, the solution describes a smooth
bounce between two phases of Minkowski spacetime. For $|m|=m_0$,
the solution smoothly runs from a warp product of hyperbolic
$2n$-space and two-dimensional de Sitter spacetime in the infinite
past to Minkowski spacetime in the infinite future. Lastly, for
$|m|>m_0$, there are closed time-like curves.

\subsection{Odd dimensions: $D=2n+1$}

The Kerr-Schild solution in $D=2n+1$ dimensions, with a sphere of
$d=2n-1$ dimensions, has the metric \cite{mp}
%%%%
\bea ds_D^2&=& - dt^2 + \fft{\Delta\, dr^2}{\prod_{i=1}^n
H_i-\fft{2m}{r^{2n-2}}} + r^2\,
\sum_{i=1}^n H_i\, (d\mu_i^2 + \mu_i^2\, d\phi_i^2)\nn\\
&&+\fft{2m}{r^{2n-2}\, \Delta}\, (dt -\sum_{i}^n \ell_i\,
\mu_i^2\,d\phi_i)^2\,, \eea
%%%%
where
%%%
\be \sum_{i=1}^n \mu_i^2=1\,,\qquad \Delta =\Bigl( \sum_{i=1}^n
\fft{\mu_i^2}{H_i}\Bigr)\, \prod_{i=1}^n H_i\,, \ee
%%%%
and $H_i=1 + \ell_i^2/r^2$.  We can now perform the analytical
continuation
%%%
\bea && t\rightarrow {\rm i}\,z\,,\qquad r\rightarrow {\rm
i}\,t\,,\qquad \phi_n\rightarrow {\rm i}\,\phi_n\,,\qquad
\alpha \rightarrow {\rm i}\,\alpha\,,\nn\\
&& m\rightarrow -{\rm i}\,m\,,\qquad
\ell_j \rightarrow -{\rm i}\,\ell_j\,,\qquad j=1,\cdots ,n-1\,,
\eea
%%%%
and $\ell_n$ is unchanged. We define $\mu_n=\td\mu_n$ and
$\mu_j={\rm i}\, \td \mu_j$, such that $\td\mu_n^2 -
\sum_{j=1}^{n-1} \td\mu_j^2=1$. The Kerr-Schild solution becomes
an S-Kerr solution, given by
%%%%%
\bea ds_D^2&=& dz^2 - \fft{\td\Delta\, dt^2}{\prod_{i=1}^n \td
H_i-\fft{2m}{t^{2n-2}}}+ t^2\,
\sum_{i=1}^{n-1} \td H_i\, (d\td\mu_i^2 + \td\mu_i^2\, d\phi_i^2)\nn\\
&& +\td H_n\,(-d\td\mu_n^2+\td\mu_n^2\, d\phi_n^2)
-\fft{2m}{t^{2n-2}\, \td\Delta}\, (dz -\sum_{i}^n \ell_i\,
\td\mu_i^2\,d\phi_i)^2\,,\label{skerrodd} \eea
%%%%%
where
%%%
\bea \td \Delta &=& \Bigl(\td\mu_n^2 -\sum_{j=1}^{n-1}
\fft{\td\mu_j^2}{\td H_j}\Bigr) \prod_{i=1}^n\td H_i\,,\nn\\
\td H_n &=& 1-\fft{\ell_n^2}{t^2}\,,\qquad \td H_j=1
+\fft{\ell_j^2}{t^2}\,,\qquad j=1,\cdots,n-1 \,. \eea
%%%
This solution is invariant under $t\rightarrow -t$. In the
asymptotic $t\rightarrow \pm \infty$ regions, the metric describes
a $D-1$ dimensional universe with hyperbolic spatial slices
expanding or contracting at a constant rate, together with a
stable $\R$ or $S^1$ direction. For the metric to be regular, it
is necessary that $\ell_n=0$, and all the remaining $\ell_i$
non-vanishing.  This ensures that the term $t^{2n}\td \Delta$ is
positive definite.  The global structure is then largely
determined by the function
%%%%%
\be F=-2m + t^{2n-2} \prod_{i=1}^{n-1} \td H_i\,. \ee
%%%%
For small $m$, namely
%%%
\be
m<m_0=\ft12 \prod_{i=1}^{n-1} \ell_i^2\,,\qquad
\ell_n=0\,,\label{kerroldcon}
\ee
%%%
the function $F$ is positive definite and the metric is regular
everywhere, bouncing between two Minkowski spacetimes.   When
$m\ge m_0$, the function has two real roots $\pm t_0$, in between
which the metric becomes stationary.  The metric develops closed
time-like curves in this region.  This can be seen by looking at
the term $d\phi_1^2$, which is given by (\ref{tdeltagen}) with
$\td H_n=1$ and $\ell_n=0$. Thus, a regular solution emerges only
when the condition (\ref{kerroldcon}) is satisfied with $m_0$
non-vanishing.

\section{Non-singular S$p$-branes}

Single-charge $p$-branes are solutions of the Lagrangian
%%%%
\be e^{-1}\,{\cal L}_D= R-\fft12(\partial \phi)^2-\fft{1}{2\td n!}
e^{a\,\phi}\,(F_{\td n})^2\,, \ee
%%%%
where $F_{(\td n)}=dA_{(\td n-1)}$ and $a^2=4-2(\td n-1)(D-\td
n-1)/(D-2)$ \cite{duff}. This Lagrangian admits an electric
$(d-1)$-brane with $d=\td n-1$ and a magnetic $(d-1)$-brane with
$d=D-\td n-1$. Since the magnetic solution can be considered as an
electric solution of the dual $(D-\td n)$-form field strength
$F_{(D-\td n)}$, we need only explicitly consider the electric
solution. A dual parameter is given by $\td d=D-d-2$.

The rotating $p$-brane solutions \cite{tenauthor} can be
analytically continued to non-singular S$p$-branes. Since the
calculation is a straightforward extension of the previous cases,
here we simply present the resulting S$p$-brane solutions. If the
foliating hyperbolic space of the transverse space is of even
dimension, corresponding to $\td d=2n-1$, then the S$p$-brane
metric is given by
%%%%
\bea ds_D^2 &=& \td H^{-\fft{\td d}{D-2}}\,\Bigl[ \Bigl(
1-\fft{2m}{t^{\td d}\,\td \Delta}\Bigr)\,dz^2+dx_1^2+\cdots
+dx_p^2\Bigr]\nn\\ && +\td H^{\fft{d}{D-2}}\,\Bigl[
-\fft{\td\Delta\,dt^2}{\prod_{i=1}^n \td H_i-\fft{2m}{t^{\td d}}}
+t^2\,\Big( -d\td\mu_0^2+\sum_{i=1}^n (d\td\mu_i^2+\td\mu_i^2\,
d\phi_i^2)\Big)\nn\\ &&+\fft{4m\,\cos\alpha}{t^{\td d}\,\td
H\,\td\Delta}\, dz\,(\sum_{i=1}^n \ell_i\,\td\mu_i^2\,d\phi_i)
-\fft{2m}{t^{\td d}\,\td H\,\td \Delta}\,(\sum_{i=1}^n \ell_i\,
\td\mu_i^2\,d\phi_i)^2\Bigr]\,, \label{spd}
\eea
%%%%
and the dilaton $\phi$ and gauge potential $A_{(n-1)}$ are given
by
%%%%
\be e^{2\phi/a}=\td H\,,\qquad A_{(n-1)}=\fft{1-\td
H^{-1}}{\sin\alpha}\,(\cos\alpha\,dz+\sum_{i=1}^n \ell_i\,
\td\mu_i^2\,d\phi_i)\wedge d^{n-2}x\,, \ee
%%%%
where
%%%%
\be \td\Delta=\prod_{i=1}^n \td H_i\,\Bigl( \td\mu_0^2-
\sum_{i=1}^n \fft{\td\mu_i^2}{\td H_i}\Bigr)\,,\qquad \td H=
1-\fft{2m\,\sin^2 \alpha}{t^{\td d}\,\td\Delta}\,,\qquad \td H_i
=1+\fft{\ell_i^2}{t^2}\,. \ee
%%%%
The $\td\mu_i$ satisfy $\td\mu_0^2-\sum_{i=1}^n \td\mu_i^2$. This
solution is invariant under $t\rightarrow -t$, $m\rightarrow
(-1)^{\td d}\,m$.

Since the analysis of the regularity and geometry are similar to
that of the previous solutions, we will not repeat it here. The
main point is that there is a finite $m_0$ such that, for $m<m_0$
(more precisely, $|m|<m_0$ in the case of odd $\td d$), the
geometry is regular and exhibits a smooth bounce between two
phases of Minkowski spacetime. For $m=m_0$, the geometry runs from
a warped product of two-dimensional de Sitter spacetime,
$p$-dimensional Euclidean space and a hyperbolic $2n$-space in the
infinite past to Minkowski spacetime in the infinite future. The
warp factor depends on the hyperbolic coordinates. Finally, for
$m>m_0$, there are closed time-like curves and, for $m\ge
m_0/\sin^2\alpha$, a curvature singularity.

If the foliating hyperbolic space of the transverse space is of
odd dimension, then the solution is given by (\ref{spd}) with $\td d=2n$
and $\td \mu_0=0$.  The singularity structure of the geometry is
similar to above except that, instead of having a de Sitter
component at $m=m_0$, there are closed time-like curves.

\section{Conclusions}

We have found twisted S$p$-brane and S-Kerr solutions by
analytically continuing rotating $p$-branes and higher-dimensional
Kerr black holes, respectively. There is always a region in
parameter space for which these time-dependent solutions are
completely regular, even though the corresponding static solutions
are marred by a naked singularity. This is among a growing number
of examples for which singularities of static black hole or
$p$-brane solutions can be analytically continued to extend onto a
smooth manifold.

The precise structure of these time-dependent solutions depends on
whether the foliating hyperbolic space of the transverse space is
odd or even-dimensional. In the former case, if a single $\ell_i$
vanishes, then for the range $m<m_0$ there is a smooth bounce
between two phases of Minkowski spacetime. This case includes
twisted SM2 and SD3-branes. On the other hand, for an
even-dimensional foliating hyperbolic space, a smooth bounce
between two phases of Minkowski spacetime is exhibited within the
range $|m|<m_0$, provided that none of the $\ell_i$ vanish. Also,
for $|m|=m_0$, the solution smoothly runs from a warped product
containing two-dimensional de Sitter spacetime, for which the warp
factor depends on hyperbolic coordinates, to Minkowski spacetime.
This latter case includes the twisted SM5-brane. Regardless of the
dimensionality of the foliating hyperbolic space, for $m\ge m_0$
there are closed time-like curves. In the presence of a non-zero
brane charge parameter $\alpha$, there is a curvature singularity
for $m\ge m_0/\sin^2\alpha$.

Although some of these solutions exhibit a two-dimensional de
Sitter phase, it would naturally be quite nice to find solutions
for which the spacetime element which undergoes exponential
expansion has four dimensions. Therefore, it is of substantial
interest to uncover a greater array of regular time-dependent
solutions, and taking the analytical continuation of static
solutions provides a guiding light in this endeavor.

The time evolution of our solutions may be used to investigate
particle production in the early universe, the geometrical
backreaction of tachyon condensation, and may provide regular
supergravity backgrounds on which to study the large $N$
open-closed string duality in a time-dependent context.

\section*{Acknowledgment}

J.F.V.P. is grateful to the George P. and Cynthia W. Mitchell
Institute for Fundamental Physics for hospitality during the
course of this work. We would like to thank Gary Gibbons, Don Page
and Chris Pope for useful discussions.

\end{document}